\documentclass[12pt]{iopart}

\usepackage{xspace}
\usepackage{graphicx}
\usepackage{geometry}
\usepackage{lineno}
\usepackage{cite} 
\expandafter\let\csname equation*\endcsname\relax
\expandafter\let\csname endequation*\endcsname\relax
\usepackage{amsmath}

\newcommand{\pp}    {pp\xspace}
\newcommand{\PbPb}  {\mbox{Pb--Pb}\xspace}
\newcommand{\pPb}   {\mbox{p--Pb}\xspace}

\newcommand{\Pt}           {\ensuremath{p_\mathrm{T}}\xspace}

\newcommand{\seven}         {$\sqrt{s}~=~7$~Te\kern-.1emV\xspace}
\newcommand{\fivenn}        {$\sqrt{s_{\mathrm{NN}}}~=~5.02$~Te\kern-.1emV\xspace}
\newcommand{\GeVc}          {Ge\kern-.1emV/$c$\xspace}
\newcommand{\TPC}    {\rm{TPC}\xspace}
\newcommand{\pip}    {\ensuremath{\pi^{+}}\xspace}
\newcommand{\pim}    {\ensuremath{\pi^{-}}\xspace}

\newcommand{\pbar}   {\ensuremath{\rm\overline{p}}\xspace}
\newcommand{\kzero}  {\ensuremath{{\rm K}^{0}_{\rm{S}}}\xspace}

\newcommand{\lmb}    {\ensuremath{\Lambda}\xspace}
\newcommand{\almb}   {\ensuremath{\overline{\Lambda}}\xspace}
\newcommand{\Om}     {\ensuremath{\Omega^-}\xspace}
\newcommand{\Mo}     {\ensuremath{\overline{\Omega}^+}\xspace}
\newcommand{\X}      {\ensuremath{\Xi^-}\xspace}
\newcommand{\Ix}     {\ensuremath{\overline{\Xi}^+}\xspace}
\newcommand{\Xis}    {\ensuremath{\Xi^{\pm}}\xspace}
\newcommand{\Oms}    {\ensuremath{\Omega^{\pm}}\xspace}

\newcommand{\pT}{\ensuremath{\Pt}}
\newcommand{\pTj}{\ensuremath{p_{\rm T,jet}}}
\newcommand{\pTjch}{\ensuremath{\pTj^{\rm ch}}}

\newcommand{\rhobkg}{\ensuremath{\rho_{\rm bkg}}}
\newcommand{\kT}{\ensuremath{k_{\rm T}}}
\newcommand{\akT}{anti-$\kT$\xspace}
\newcommand{\thirteen} {$\sqrt{s}~=~13$~Te\kern-.1emV\xspace}

\newcommand{\Vzero}{\ensuremath{{\rm V}^{0}}}

\newcommand{\abs}[1]{\ensuremath{\left|#1\right|}}

\begin{document}
\title[Strange particle production in jets and the UE]{Production of strange hadrons in jets and underlying events in \pp and \pPb collisions with ALICE}

\author{Pengyao Cui for the ALICE Collaboration}

\address{Central China Normal University, Wuhan, 430079, P. R. China}
\ead{pengyao.cui@cern.ch}
\vspace{10pt}
\begin{indented}
\item[]January 2022
\end{indented}

\begin{abstract}
The production of strange~(\kzero,  \lmb) and multi-strange~(\Xis and \Oms) hadrons in jets and underlying events in \pp and \pPb collisions is studied with ALICE at the LHC.
Transverse momentum ($\pT$) differential density distribution of particles produced in a jet is compared to that of inclusive particle production and that in underlying events.
The particle yield ratios of $(\lmb + \almb)/2\kzero$ and $(\X + \Ix)/(\lmb + \almb)$ as a function of $\pT$ are also investigated in jets and underlying events.
The production of the multi-strange hadrons, $\Xis$ and $\Oms$, and the corresponding ratio in jets and underlying events are measured for the first time.
The  $\pT$-differential density distribution of hadrons associated with hard scattering decreases slower than that for inclusive production.
The baryon-to-meson and baryon-to-baryon ratios measured in jets exhibit clear differences from values obtained from the inclusive spectra in the intermediate $\pT$ range.
The $\pT$ distribution of strange hadrons produced in jets is also studied in pp collisions at \thirteen and \seven and in \pPb collisions at \fivenn.
The $\pT$ spectra of particle produced in jets are independent of collision systems and collision energies.
No apparent collision energy and collision system dependence  is observed for the particle yield ratios of $(\lmb + \almb)/2\kzero$ and $(\X + \Ix)/(\lmb + \almb)$ associated with energetic jets.
These new results are compared to PYTHIA 8 event generator. The $(\lmb + \almb)/2\kzero$ ratio are generally reproduced by the model, but large discrepancies between data and PYTHIA simulations are observed for $(\X + \Ix)/(\lmb + \almb)$.
\end{abstract}

%
\vspace{2pc}

\section{Introduction}\label{sec:intr}
The unprecedented energies available at the Large Hadron Collider (LHC) provide unique opportunities to investigate the properties of the quark-gluon plasma (QGP)~\cite{Rafelski:1980rk, Satz:2000bn, Shuryak:1983ni, Jacak:2012dx, Cleymans:1985wb, Bass:1998vz, BraunMunzinger:2007zz}.
Recent reports on the enhancement of (multi-)strange hadrons~\cite{Abelev:2013haa, ALICE:2017jyt, Khachatryan:2016yru, ALICE:2016fzo, ALICE:2015mpp}, double-ridge structure~\cite{Aad:2015gqa, Abelev:2012ola, ABELEV:2013wsa, CMS:2012qk, ALICE:2012eyl}, non-zero $v_{2}$ coefficients~\cite{Acharya:2019vdf, Khachatryan:2015waa, ALICE:2014dwt}, mass ordering in hadron $\pT$ spectra~\cite{ALICE:2013wgn}, and characteristic modifications of baryon-to-meson ratios~\cite{ALICE:2017jyt, Khachatryan:2016yru, Acharya:2018orn, Abelev:2013xaa} suggest that collective phenomena are present at the LHC energies also in \pp and \pPb collisions.
Those results indicate that the collective effects are not unique to heavy-ion collisions.
However, several measurements show the absence of a nuclear effect on the jet production at mid-rapidity in \pp and \pPb collisions~\cite{ALICE:2017svf, Acharya:2019jyg, Acharya:2019tku, ALICE:2014dla, Abelev:2013fn, Acharya:2018eat, Acharya:2017okq, Adam:2015xea, Adam:2016jfp}. Therefore, it is of great interest to investigate the origin of the collectivity-driven features in small systems (\pp and \pPb).

The baryon-to-meson yield ratio as a function of $\pT$ exhibits an enhancement at intermediate $\pT$ (around 3~\GeVc) in high multiplicity small collision systems with respect to that in low multiplicity, which is qualitatively similar to that observed in \PbPb collisions~\cite{ALICE:2018pal}.
The particles with a larger strangeness content are observed to be produced more abundantly for high multiplicity in \pp collisions~\cite{ALICE:2016fzo, ALICE:2019avo} and in \pPb collisions~\cite{ALICE:2013wgn, ALICE:2015mpp}.
These effects suggest the existence of a common underlying mechanism determining the chemical composition of particles produced in these three collision systems.
The particle production at LHC energies has both soft and hard-scattering origins.
To understand the strange particle production mechanisms in small collision systems, separation of particle produced in hard processes (hard scattering) from those of the underlying event (UE) is important.

In this proceeding, the production of strange and multi-strange hadrons in jets and UE in \pp and \pPb collisions will be investigated with ALICE at LHC. The production of strange particles is studied separately within the jet and the UE region which separates the contribution associated with hard and soft processes.
The hard process is tagged by selecting a reconstructed charged-particle jets with transverse momentum $\pTj > 10$~\GeVc using the \akT algorithm~\cite{Cacciari:2008gp} with a resolution parameter $R = 0.4$. Particle production in UE is estimated in the perpendicular cone (PC) to the jet axis with a radius $R_\mathrm{cone} = 0.4$. The inclusive production of hadrons in minimum bias events is compared to that in jets and in underlying events.
The ratio of $(\lmb + \almb)/2\kzero$ and $(\X + \Ix)/(\lmb + \almb)$ as a functions of particle transverse momentum is presented in charged-particle jets and UE. The result is compared to the PYTHIA 8 simulations with the color rope model~\cite{Sjostrand:2014zea, Biro:1984cf, Bialas:1984ye}. The $(\X + \Ix)/2\kzero$ and $(\Om + \Mo)/2\kzero$ ratios, which carry the information about multi-strange particles, are also investigated by the PYTHIA~8 simulation.

\section{Experimental setup and analysis strategy}\label{sec:ana}
The data samples used in this analysis were recorded by the ALICE detector~\cite{ALICE:2008ngc, ALICE:2014sbx} during LHC \pp run in 2010 (\seven), 2015 (\thirteen) and in \pPb runs in 2013 (\fivenn).
A detailed description of the ALICE apparatus and its performance can be found in~\cite{Collaboration_2008, Abelev:2014ffa}.
This analysis relies on the central tracking system and the forward V0 detector~\cite{Abbas:2013taa} in ALICE. 
The Inner Tracking System~(ITS)~\cite{Aamodt:2010aa} covering pseudorapidity of $|\eta| < 0.9$ is used to measure the charged particle and to provide measurement of the primary interaction vertex (PV). 
The Time Projection Chamber~(TPC)~\cite{Alme:2010ke}, also one of the central barrel detectors, is the main detector used for charged particle measurement and covers the pseudo rapidity range $|\eta| < 0.9$.  The TPC provides in addition information about particle type obtained through measurements of d$E$/d$x$. The two forward scintillator arrays V0A (covering pseudo-rapidity range of $2.8 < \eta < 5.1$), and V0C ($-3.7 < \eta < -1.7$) combined with ITS are employed for providing the event trigger information.

The strange particles \kzero, \lmb, \almb, and \Xis are reconstructed at mid-pseudorapidity ($\abs{\eta} < 0.75$) via their specific weak decay topology.
The following charged decay channels are used~\cite{PhysRevD.98.030001}: 
$$
\begin{aligned}
\kzero      & \to \pip + \pim               & B.R. & = (69.20 \pm 0.05) \%, \\
\lmb (\almb) & \to \mathrm{p}(\pbar) + \pim (\pip)    & B.R. & = (63.9  \pm 0.5)  \%, \\
\X (\Ix)    & \to \lmb (\almb) + \pim (\pip) & B.R. & = (99.887 \pm 0.035) \%. \\
\end{aligned}
$$
The proton and pion tracks are identified in the TPC via their measured energy deposition~\cite{Abelev:2014ffa}.
The identification method of the \Vzero (\kzero and \lmb (\almb) which decays into two oppositely charged daughter particles) and \Xis candidate is the same as in earlier ALICE publication~\cite{Aamodt:2011zza,Abelev:2012jp,Acharya:2018orn,Abelev:2013haa,ALICE:2020jsh,Acharya:2019kyh}.
The signal extraction is performed as a function of $\pT$.
In each $\pT$ interval, an invariant mass histogram is filled with the corresponding counts. The raw yield (signal number) of strange hadrons in each $\pT$ intervals is extracted by bin counting method~\cite{ALICE:2020jsh}.

Charged-particle jets are reconstructed with the \akT algorithm~\cite{Cacciari:2008gp} in the FastJet package\cite{Cacciari:2011ma}, with a resolution parameter $R = 0.4$. 
The charged particle tracks, which are used as input for jet reconstruction, are selected in $\abs{\eta_{\rm trk}} < 0.9$ ($\TPC$ acceptance). The transverse momentum ($\pT$) of tracks should be larger than 0.15~\GeVc. Pseudorapidity of the charged-particle jet is constrained to  $|\eta_\mathrm{jet}| < 0.35~( = 0.75 - 0.4)$. These conditions ensure that the jet cone is fully overlapping with the acceptances of both charged-particle tracks and strange particles ($|\eta| < 0.75$). The transverse-momentum density of the background ($\rhobkg$), originating from the UE and/or pile-up, contributes to the jet energy reconstructed by the jet finder. The background density~($\rhobkg$) determined by the $\kT$ algorithm~\cite{Catani:1993hr, Ellis:1993tq} in \pp collisions is negligible and not subtracted. In \pPb collisions, an estimator adequate for the more sparse environment than \PbPb collisions is employed by scaling $\rhobkg$ with an additional factor to account for event regions without particles~\cite{ALICE:2015umm}. A transverse momentum cut on the charged-particle jets, $\pTj^{\rm ch} > 10$~\GeVc, is applied to tag the hard scattering processes~\cite{Acharya:2021oaa}.

The strategy of obtaining strange hadrons in charged-particle jets, follows that presented in~\cite{Acharya:2021oaa}. The distance of particle to the jet axis on the $\eta - \varphi$ plane is defined as: $$
	R(\mathrm{par, ~jet}) = \sqrt{\left( \eta_\mathrm{jet} - \eta_\mathrm{par} \right)^{2} + \left( \varphi_\mathrm{jet} - \varphi_\mathrm{par} \right)^{2}}.
$$
When the distance of particle to jet axis $R(\mathrm{par, ~jet})$ is less than the jet cone size $R_\mathrm{cone} = 0.4$, the particle is considered inside the jet cone (JC).
The remaining contribution from the underlying event (UE) in the JC selection, which refers to the particle not associated with jet fragmentation, is estimated in the perpendicular cone (PC) to the jet axis with radius $R_{\rm PC} = 0.4$. The systematic uncertainty due to the UE subtraction is estimated by using two methods:
\begin{itemize}
    \item \textbf{Outside cone (OC):} particles are reconstructed outside the jet cone of the reconstructed jet in the event, i.e. $R(\mathrm{par, ~jet}) > R_\mathrm{cone}$.
    \item \textbf{Non-jet events (NJ):} particles found in events without any jet with $\pT > 5$~\GeVc
\end{itemize}

To obtain the corrected spectra, the acceptance and efficiency factors as a function of $\pT$ have to be computed. The acceptance and efficiency of each particle are obtained from Monte Carlo simulated data. Due to differences in the experimental acceptance for particles associated with jets and the UE, efficiencies of particles are estimated separately for each case~\cite{Acharya:2021oaa}.

The per event $p_\mathrm{T}$-differential density $\mathrm{d} \rho/ \mathrm{d} p_\mathrm{T}$ of strange particles is defined as: 
$$
	\frac{\mathrm{d} \rho}{\mathrm{d} p_\mathrm{T}} = \frac{1}{N_\mathrm{event}} \times \frac{1}{\langle \mathrm{Area} \rangle } \times \frac{\mathrm{d} N}{\mathrm{d} p_\mathrm{T}},
$$
where $N_\mathrm{event}$ is the number of events, $\langle \mathrm{Area} \rangle $ is the acceptance area in pseudorapidity and azimuthal angle ($\Delta \eta \times \Delta \varphi$), and $\frac{\mathrm{d} N}{\mathrm{d} p_\mathrm{T}}$ is the strange particle yield in each $p_\mathrm{T}$ bin.
This density is evaluated for particles in the jet cone (JC) and UE.
The density of particles coming from jet fragmentation (JE) is calculated by
$$	\frac{\mathrm{d} \rho_{\rm JE}}{\mathrm{d} p_\mathrm{T}} = \frac{\mathrm{d} \rho_{\rm JC}}{\mathrm{d} p_\mathrm{T}} - \frac{\mathrm{d} \rho_{\rm UE}}{\mathrm{d} p_\mathrm{T}}.$$


\section{Results}\label{sec:res}

\begin{figure}[ht]
	\begin{center}
		\includegraphics[width=.49\textwidth]{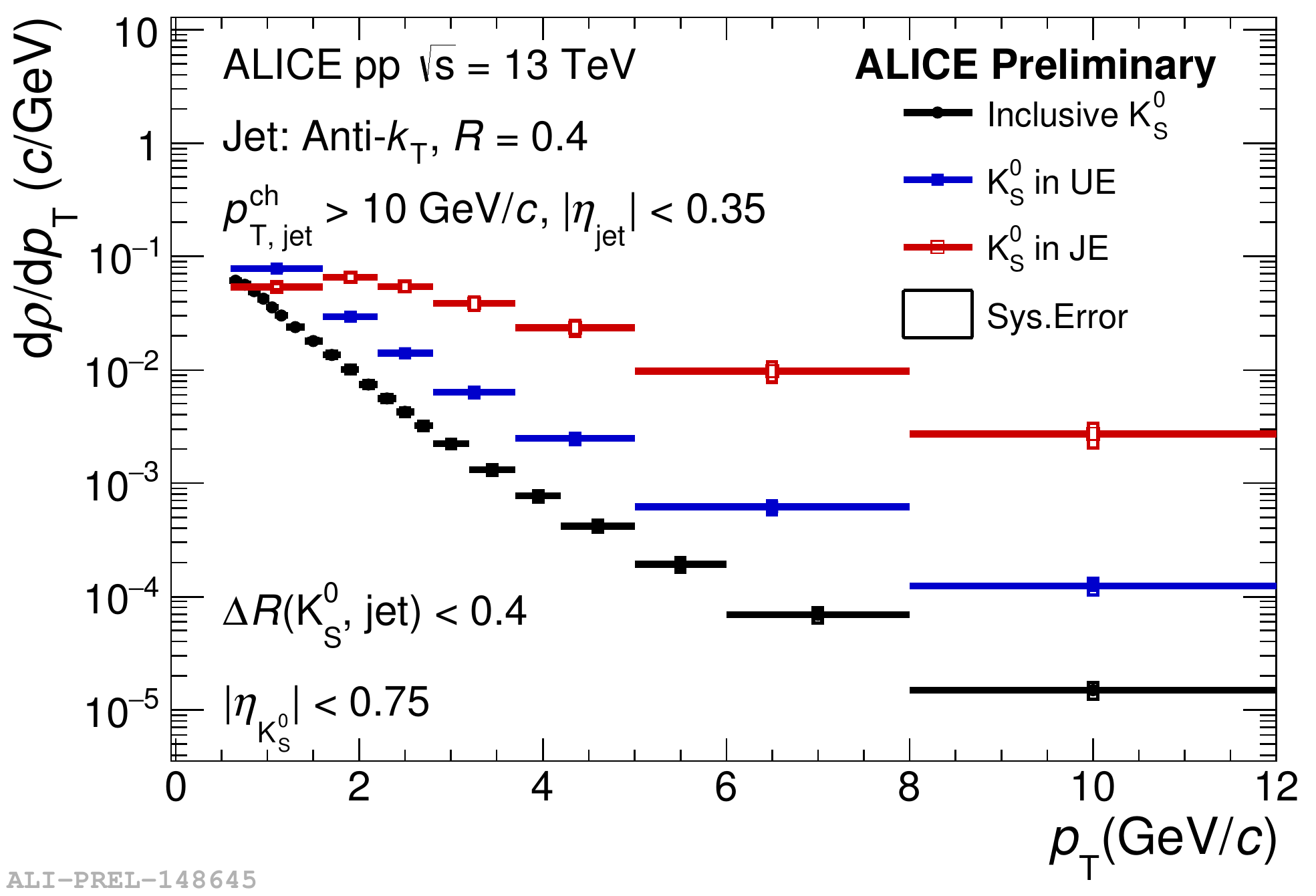}\\
		\includegraphics[width=.49\textwidth]{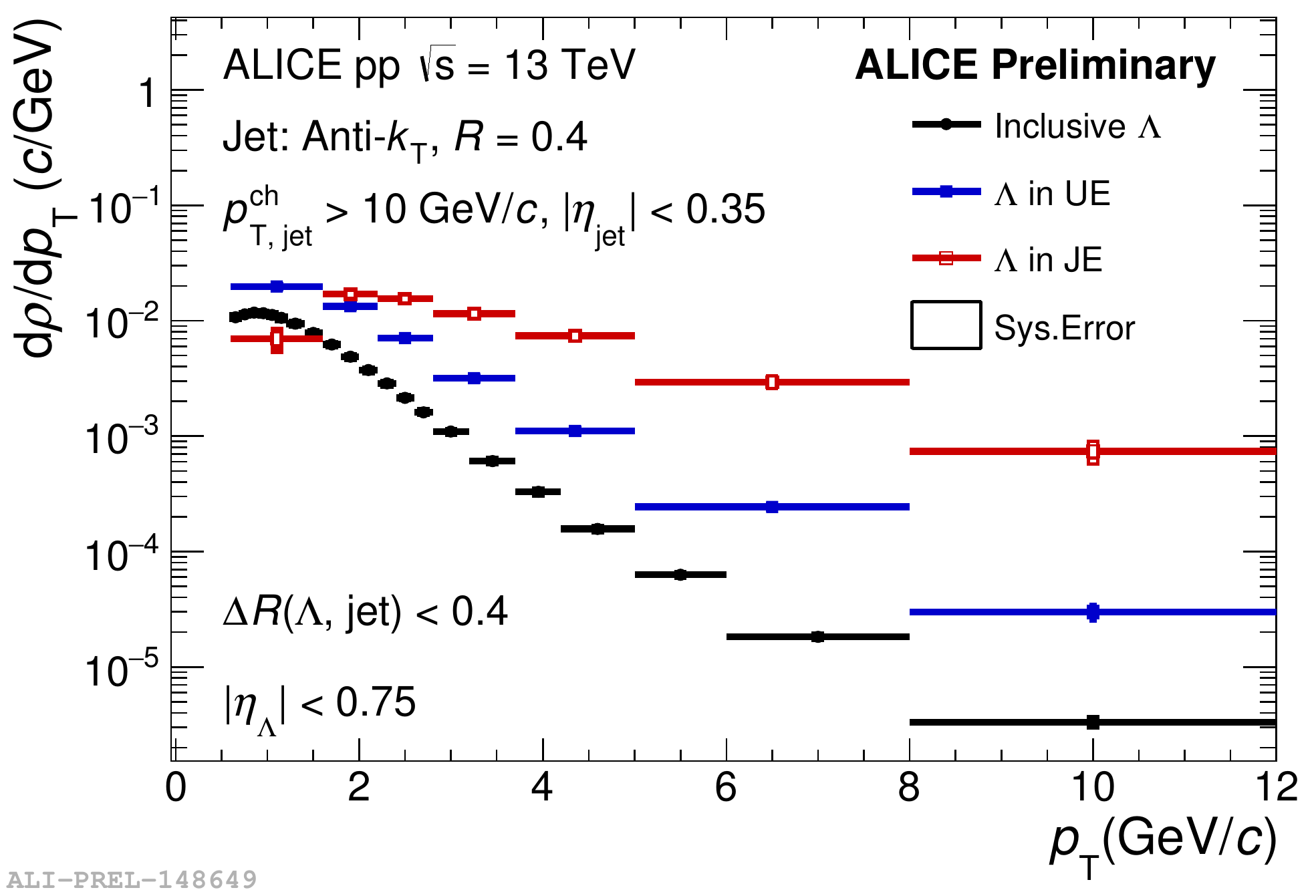}
		\includegraphics[width=.49\textwidth]{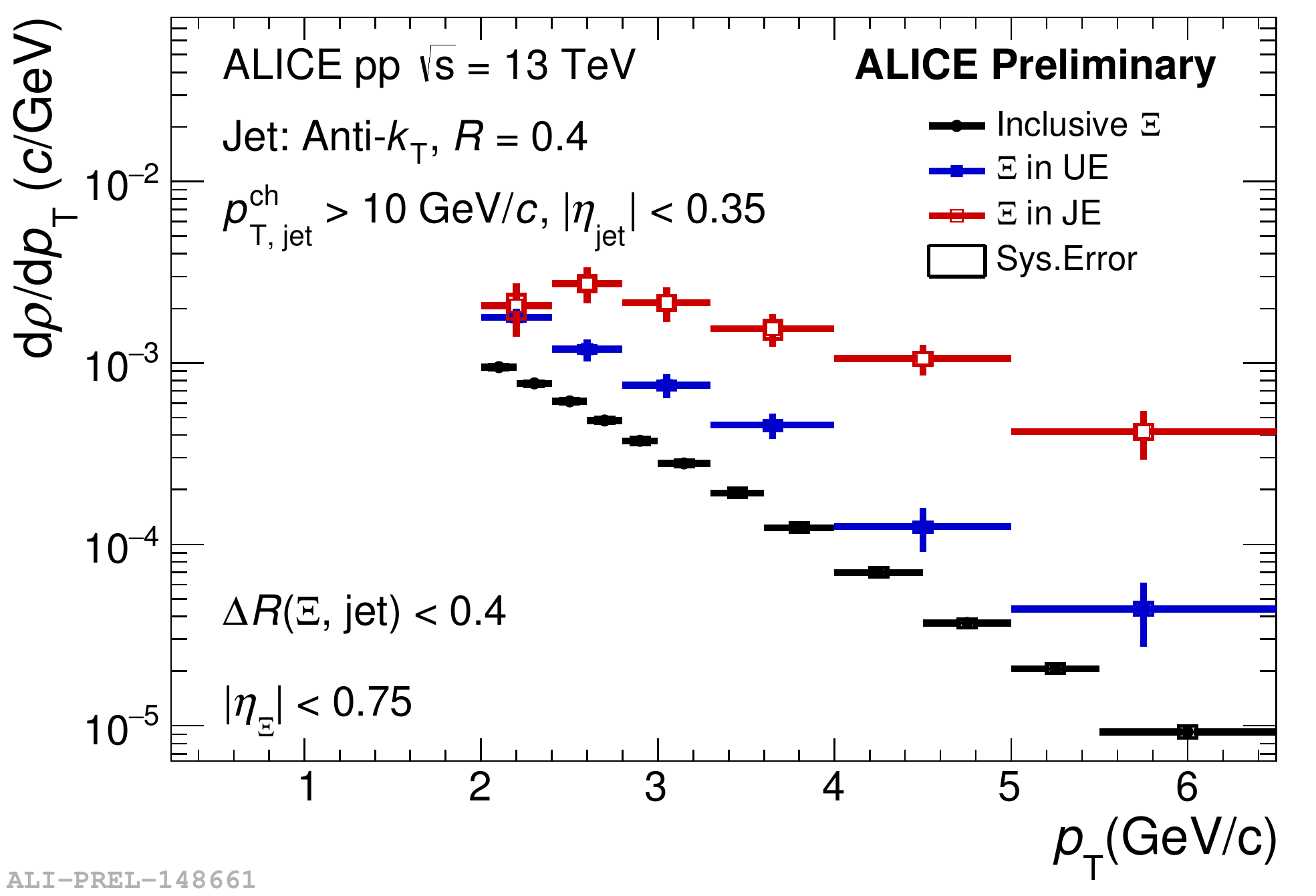}
	\end{center}
	\caption{(color online) $p_\mathrm{T}$-differential density of $\mathrm{K^0_S}$ meson (top), \lmb (bottom$-$left) and $\Xi^{-} + \overline{\Xi}^{+}$ (bottom$-$right) baryons in \pp collisions at $\sqrt{s} = 13$~TeV. Black points correspond to particles from minimum-bias events, blue points correspond to particles within a cone perpendicular to the jet axis, which are associated with the UE, and red points correspond to the particles from the jet fragmentation.}
	\label{fig:spectra}
\end{figure}

The result of \kzero, \lmb, \Xis and \Oms productions in \pp collisions at \seven, \thirteen and in \pPb collisions at \fivenn are presented in this section.
The fully corrected $\pT$-differential densities of $\mathrm{K^0_S}$, \lmb and $\Xi^{-} + \overline{\Xi}^{+}$ obtained with various selections in \pp collisions at \thirteen are shown in figure~\ref{fig:spectra}. The inclusive particle spectra are shown in the black solid circles. The particle spectra in charged-particle jets (labeled as ``JE") and in the underlying event (labeled as ``UE") are presented with red and blue solid squares, respectively. 
As expected, the density distributions in JE, which are corresponding to jet fragmentation, are considerably harder than that in UE. This is consistent with the expectation that high-$\pT$ particles originate from the fragmentation. The density distributions of the inclusive particles obtained in minimum-bias events are also compared with that of the JE and UE particles in figure~\ref{fig:spectra}. Due to trigger bias, particle density in UE containing at least one jet with $\pTjch > 10$~\GeVc is higher than that of inclusive particles obtained in minimum-bias events.

\begin{figure}[ht]
	\begin{center}
		\includegraphics[width=.49\textwidth]{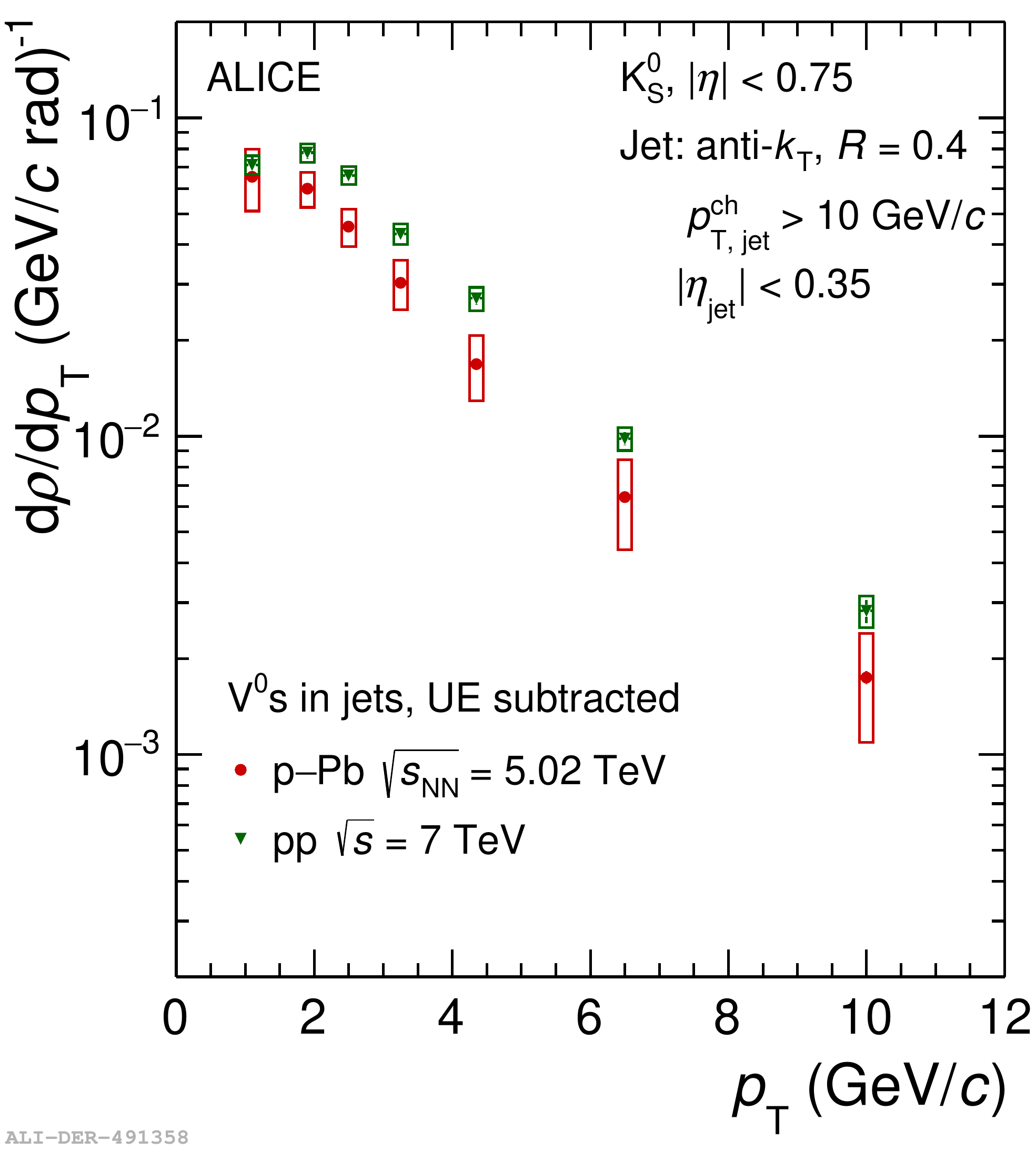}
		\includegraphics[width=.49\textwidth]{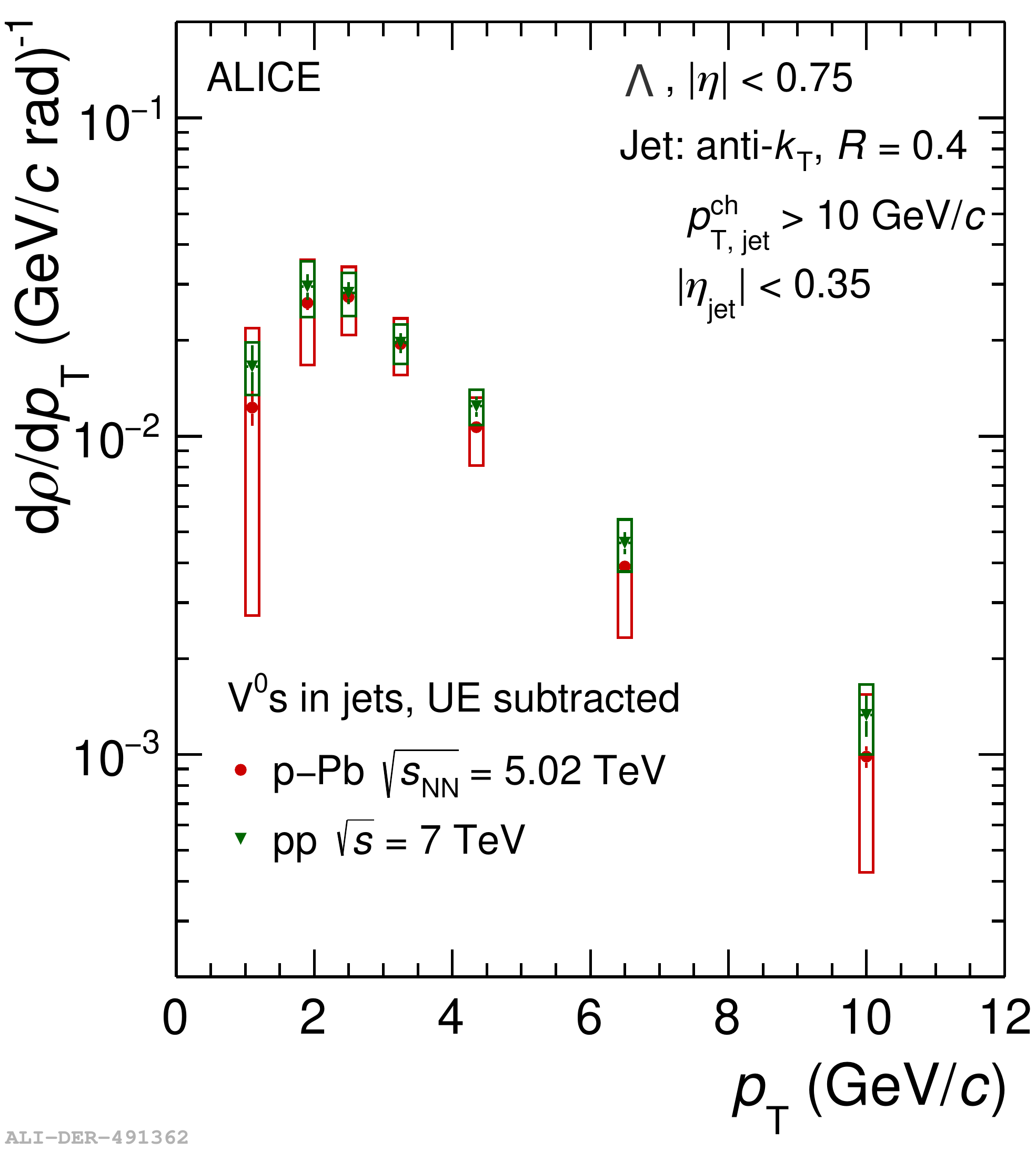}
	\end{center}
	\caption{(color online) $p_\mathrm{T}$-differential density of $\mathrm{K^0_S}$ (left) and $\lmb + \almb$ (right) in charged-particle jets in \pp collisions at \seven and in \pPb collisions at \fivenn.}
	\label{fig:comp_pp_ppb}
\end{figure}

Figure~\ref{fig:comp_pp_ppb} shows the comparison of the $p_\mathrm{T}$-differential density of $\mathrm{K^0_S}$ (left) and $\lmb$ (right) produced in charged-particle jets in \pp collisions at \seven and in \pPb collisions at \fivenn, respectively~\cite{ALICE:2021cvd}. The results indicates a weak dependence of collision system and system energy of the $\pT$-differential particle density in jets.

\begin{figure}[tbh]
	\begin{center}
		\includegraphics[width=.7\textwidth]{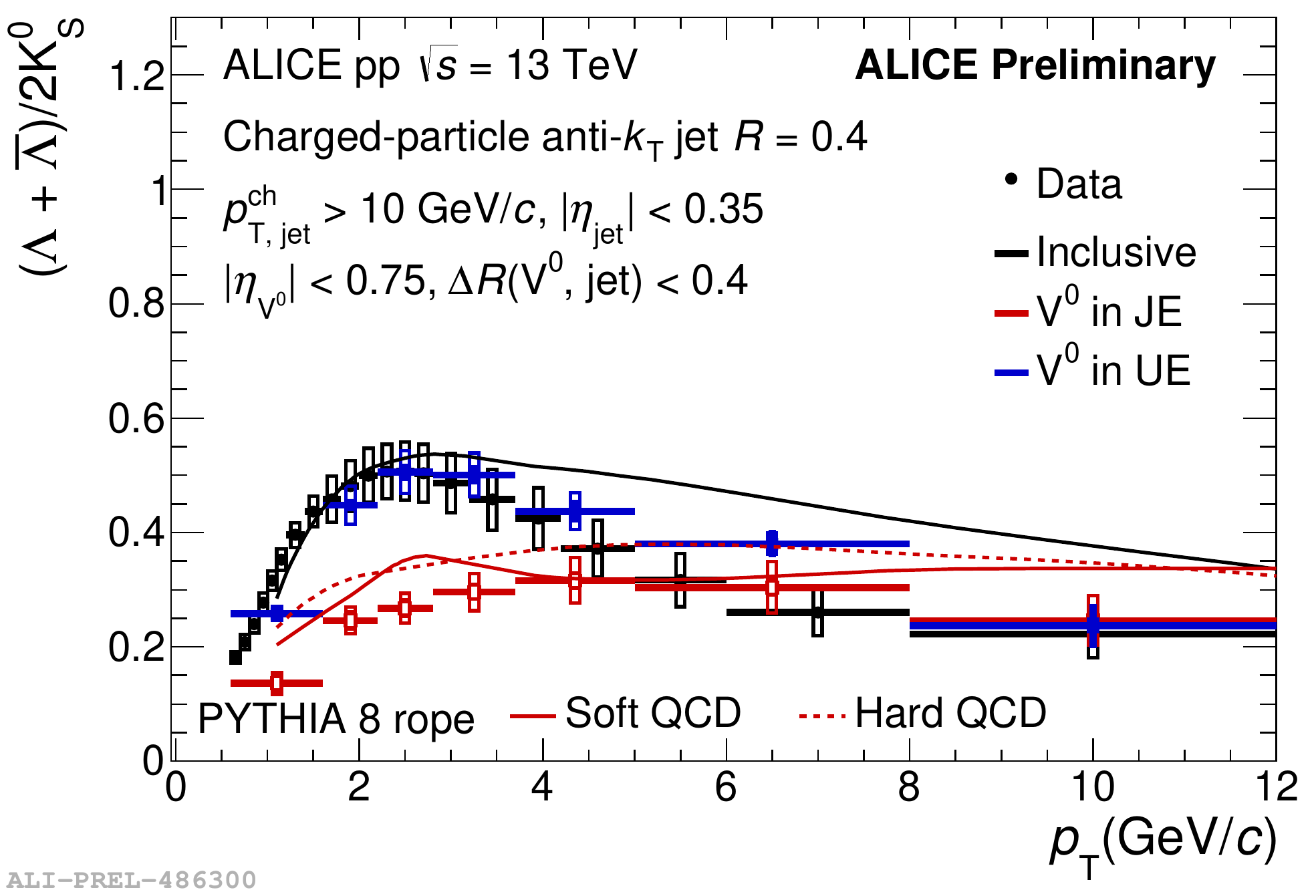}
	\end{center}
	\caption{(color online) The $(\lmb + \almb)/2\kzero$ ratio as a function of particle $p_{\rm T}$ in pp collisions at $\sqrt{s}$ = 13~TeV. Black points correspond to the ratio evaluated for particles from minimum bias events, blue points correspond to the ratio obtained for particles which come from a cone that is perpendicular to the jet, and red points correspond to the ratio from the jet fragmentation.
	The solid and dashed lines are correspond to the corresponding simulation results with PYTHIA 8 color rope model.}
	\label{fig:LKratios}
\end{figure}
The $(\Lambda + \overline{\Lambda})/2\mathrm{K^0_S}$ ratios with different selection criteria as the function of $p_\mathrm{T}$ in \pp collisions at \thirteen are presented in figure~\ref{fig:LKratios}. The result suggests that the inclusive and the UE ratios manifest an enhancement at $p_{\rm T}$ around 2$-$3~\GeVc. The ratios of JE particles are significantly lower than the inclusive and UE case at low and intermediate $p_{\rm T}$. The $(\lmb + \almb)/2\kzero$ ratio in charged-particle jets is approximately independent on $\pT$ in the region beyond 3~\GeVc, especially without a maximum at intermediate $\pT$. This suggests that the enhancement of the baryon-to-meson ratio at intermediate $p_{\rm T}$ is not driven by the jet fragmentation. In addition, the ratio of inclusive particles becomes consistent with that of JE particles for $\pT > 6$~\GeVc as the high-$\pT$ particles originate from jet fragmentation.
In figure~\ref{fig:LKratios}, the $(\lmb + \almb)/2\kzero$ ratios are compared with the PYTHIA 8 color rope simulation~\cite{Sjostrand:2014zea}. It is shown that the PYTHIA 8 can generally describe the trend both for inclusive and JE.

\begin{figure}[tbh]
	\begin{center}
		\includegraphics[width=.7\textwidth]{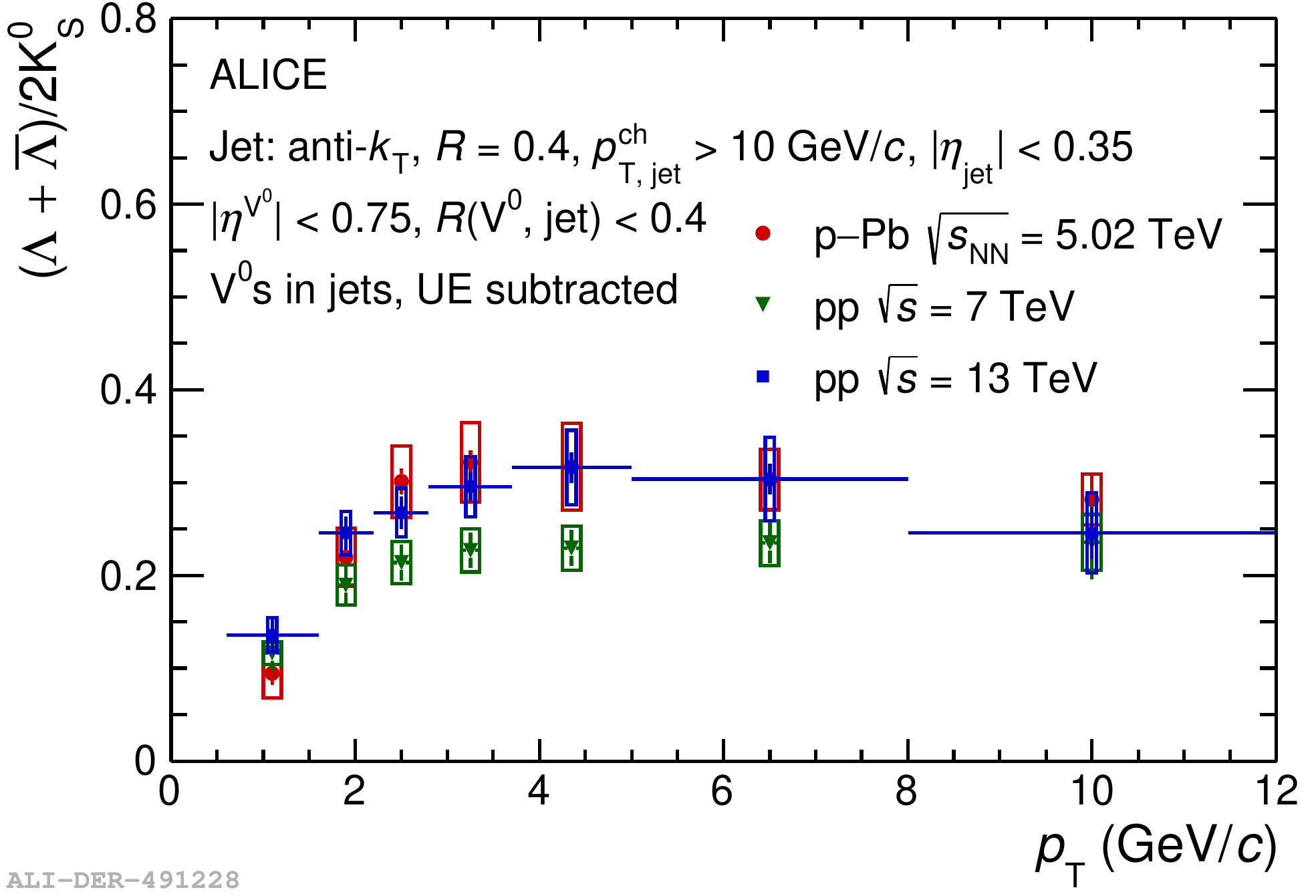}
	\end{center}
	\caption{(color online) The $(\lmb + \almb)/2\kzero$ ratio in energetic jets as a function of $p_\mathrm{T}$ in \pp collisions at \seven and \thirteen, and in \pPb collisions at \fivenn.}
	\label{fig:compLKratio}
\end{figure}

The $(\lmb + \almb)/2\kzero$ ratio in charged-particle jets in \pp collisions at \thirteen is compared to that in \pp collisions at \seven and \pPb collisions at \fivenn in figure~\ref{fig:compLKratio}~\cite{ALICE:2021cvd}. The $(\lmb + \almb)/2\kzero$ ratio in \pp collisions at \seven is lower than that in other two cases, but the deviation is less than $2\sigma$. This should be investigated further with larger statistics. Currently, no visible dependence on the collision system and collision energy is observed for $(\lmb + \almb)/2\kzero$ ratio produced by jets.

\begin{figure}[tbh]
	\begin{center}
		\includegraphics[width=.7\textwidth]{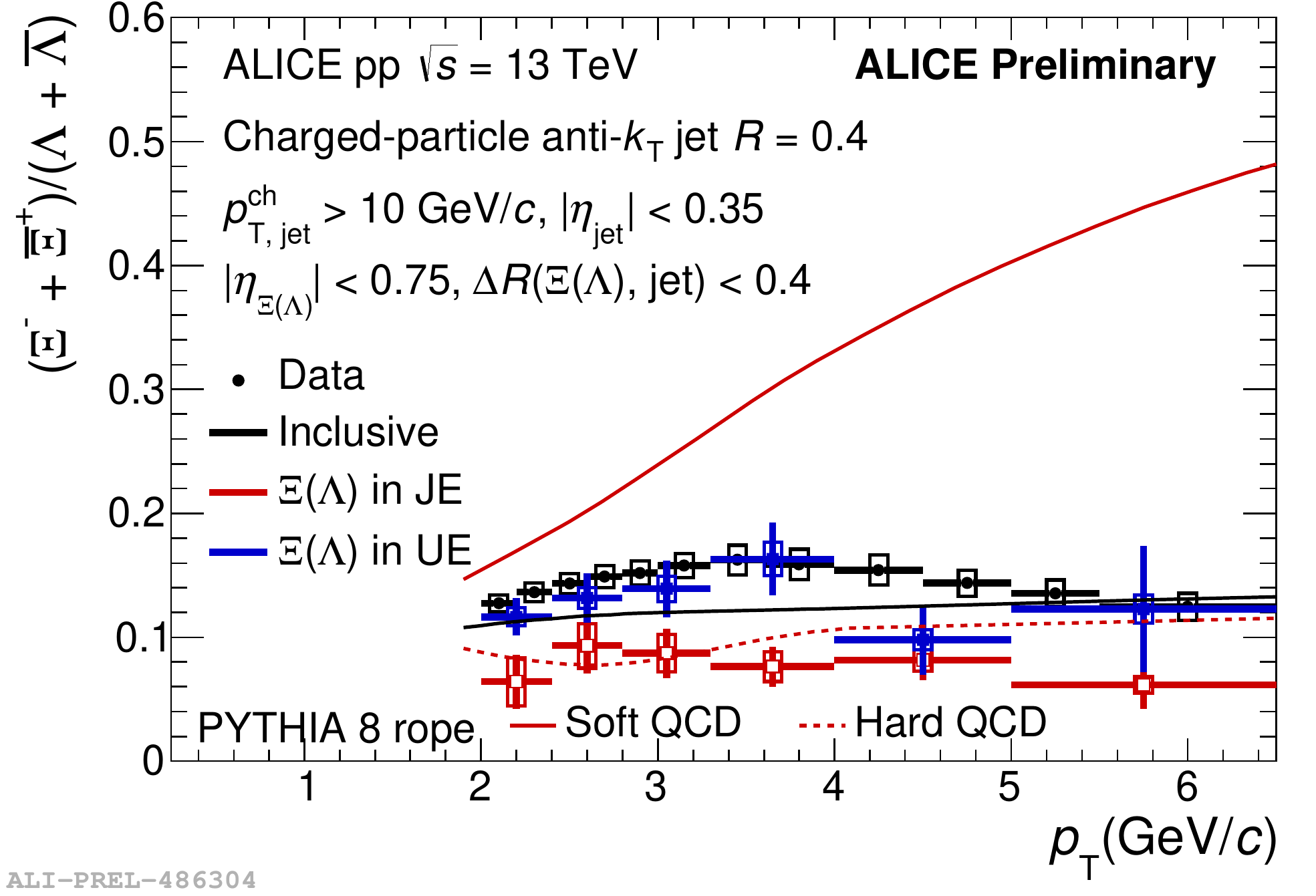}
	\end{center}
	\caption{(color online) The $(\X + \Ix)/(\lmb + \almb)$ ratio as a function of particle $p_{\rm T}$ in pp collisions at $\sqrt{s}$ = 13~TeV. Black points correspond to the ratio evaluated for particles from minimum bias events, blue points correspond to the ratio obtained for particles which come from a cone that is perpendicular to the jet, and red points represent the ratio from the jet fragmentation. The solid and dashed red lines represent the simulation results for the ratio in JE with PYTHIA 8 color rope model. The solid black line represent the simulation results for the inclusive one.}
	\label{fig:XLratios}
\end{figure}

The $(\X + \Ix)/(\lmb + \almb)$ ratio in different selections is also investigated in figure~\ref{fig:XLratios}. In this case, the numerator contains one more strange quark than the denominator. Similar to the $(\lmb + \almb)/2\kzero$ ratio shown in figure~\ref{fig:LKratios}, the $(\X + \Ix)/(\lmb + \almb)$ ratios from UE are consistent with that from the inclusive measurement. The ratio of JE particles is systematically lower than the inclusive measurement. The ratios of inclusive and UE particles increase with $\pT$ till around 4~\GeVc. However, in the measured $\pT$ acceptance, the ratio of JE particles is almost independent on $\pT$. This implies the production mechanism of $\Xi$, as multi-strange particle, in jets may be different with that in the UE. The results are compared with the PYTHIA 8 simulations in figure~\ref{fig:XLratios}. There are large discrepancies between data and PYTHIA 8 simulations. The result in PYTHIA 8 simulation is largely underestimating the inclusive $(\X + \Ix)/(\lmb + \almb)$ ratio. The ratio of JE particles given by PYTHIA 8 soft QCD increases dramatically with $\pT$. On the contrary, the ratio of JE particles given by hard QCD is more reasonable.

\begin{figure}[ht]
	\begin{center}
		\includegraphics[width=.49\textwidth]{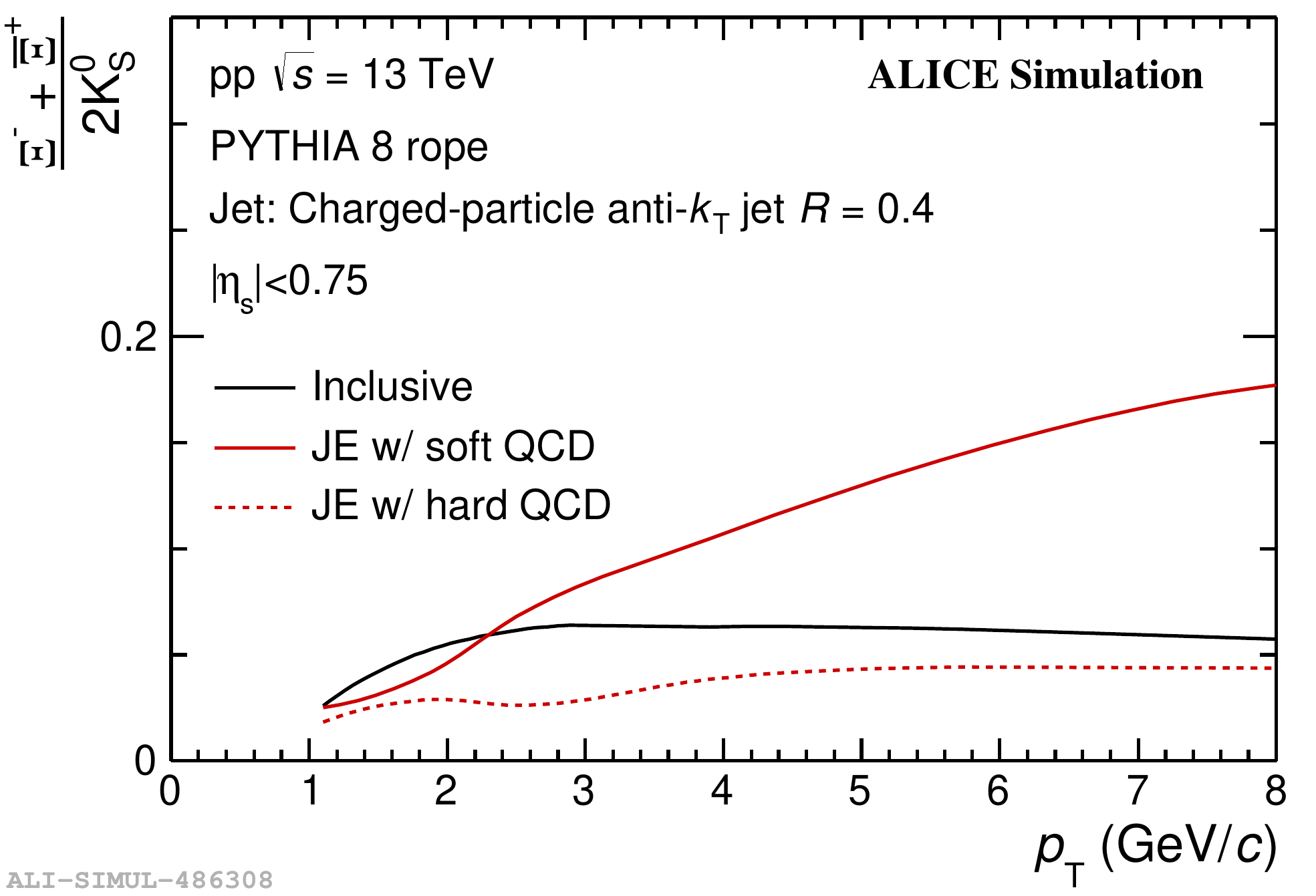}
		\includegraphics[width=.49\textwidth]{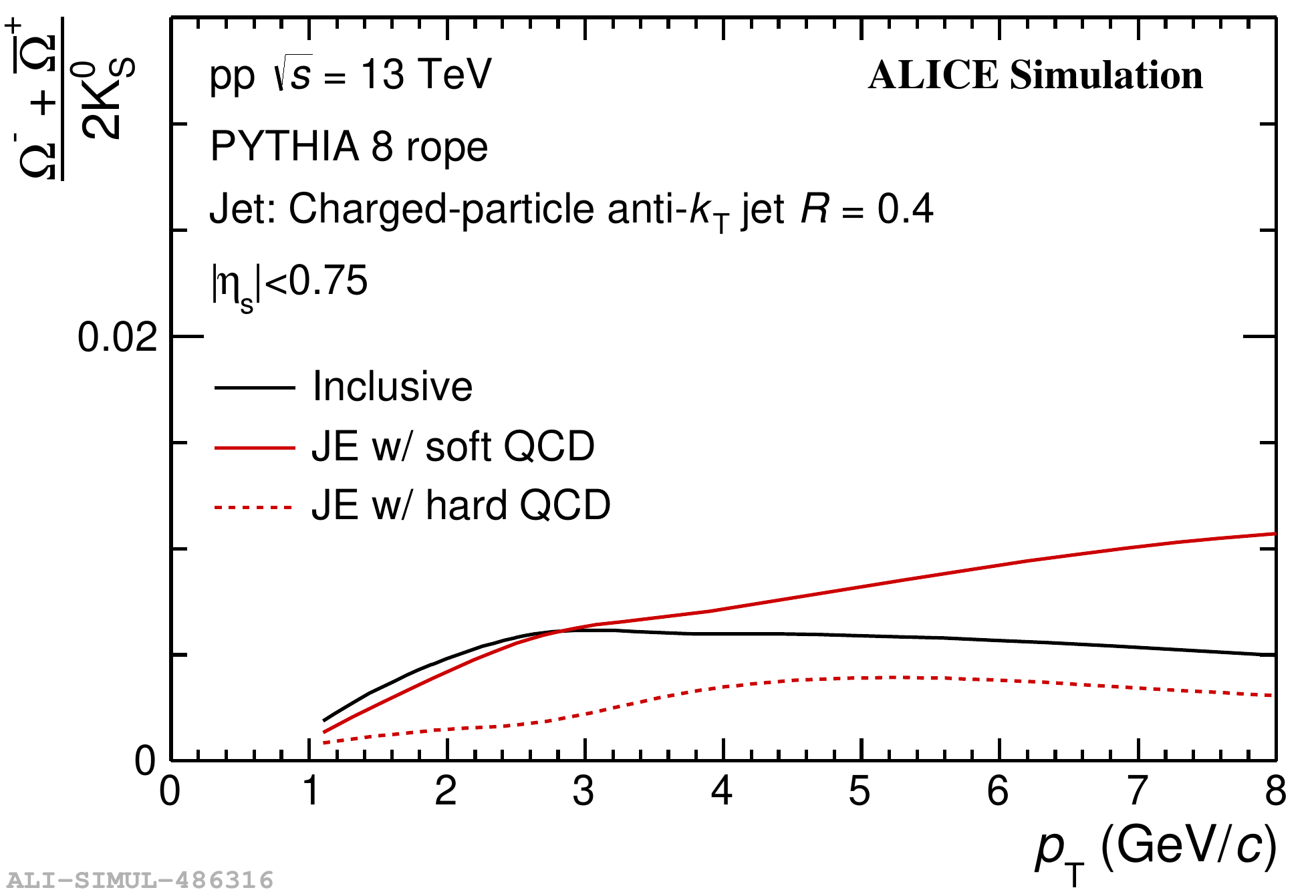}
	\end{center}
	\caption{(color online) The $(\Xi^{-} + \overline{\Xi}^{+})/2{\rm K^{0}_{S}}$ (left) and $(\Omega^{-} + \overline{\Omega}^{+})/2{\rm K^{0}_{S}}$ (right) 
	ratios obtained from PYTHIA~8 simulations.}
	\label{fig:sim}
\end{figure}

The multi-strange baryon-to-meson ratios, $(\Xi^{-} + \overline{\Xi}^{+})/2\kzero$ and $(\Om + \Mo)/2\kzero$, which carry the information about multi-strange particles, are also investigated by PYTHIA 8 simulation, shown in figure~\ref{fig:sim}.
It can be seen that PYTHIA with the soft QCD mode predicts similar strong increase as observed above. Our measurement thus provides important constraints on the production mechanisms of particles, especially for those in multi-strange sector.

\section{Summary}\label{sec:sum}

In summary, the production of \kzero, \lmb, $\Xi$ and $\Omega$ is measured separately for particles associated with hard scatterings and the UE in \pp and \pPb collisions. The baryon-to-meson and baryon-to-baryon ratios in \pp collisions at \thirteen are compared with PYTHIA 8 simulations. The $(\lmb + \almb)/2\kzero$ ratio can generally be reproduced by the model, but large discrepancies between data and PYTHIA simulations are observed for the $(\X + \Ix)/(\lmb + \almb)$ ratio. The $\pT$-differential
density of particles in jets is almost independent on collision systems and collision
energies. The baryon-to-meson and baryon-to-baryon ratios in jets are systematically lower than the inclusive measurement and independent of the particle $\pT$. The baryon-to-meson ratio enhancement has been linked to the interplay of radial flow and parton recombination at intermediate $\pT$. However, this enhancement is not seen within the jet, which indicates that these effects are indeed limited to the soft particle production processes.


\section*{Acknowledgements}
This work is supported by the National Natural Science Foundation
of China under Grant Nos. 11875143 and 12061141008.

\section*{References}
\bibliographystyle{iopart-num.bst}
\bibliography{iopart-num.bib}

\appendix

\clearpage

\end{document}